\documentclass[preprint,12pt,3p]{elsarticle}

\usepackage{array}
\newcolumntype{L}[1]{>{\raggedright\let\newline\\\arraybackslash\hspace{0pt}}p{#1}}
\newcolumntype{C}[1]{>{\centering\let\newline\\\arraybackslash\hspace{0pt}}p{#1}}
\newcolumntype{R}[1]{>{\raggedleft\let\newline\\\arraybackslash\hspace{0pt}}p{#1}}
\usepackage{tabularx}
\usepackage[bottom]{footmisc}
\usepackage{longtable}
\usepackage{threeparttablex}
\usepackage{threeparttable}
\usepackage{booktabs}
\usepackage{amssymb}
\usepackage{hyperref}
\usepackage{multirow}
\usepackage{multicol}
\usepackage{float}
\usepackage{amsthm}
\usepackage{amsmath}
\usepackage{rotating} 

\usepackage{microtype}
\usepackage{xcolor}
\journal{Travel Behaviour and Society}

\begin{document}

\begin{frontmatter}
\title{\textcolor{black}{How Infrastructure and Streetscape Shape E-Scooter Route Choice: Evidence from Washington, DC}}

\author[label1]{Yiheng Qian}
\author[label1]{Duanya Lyu}
\author[label2]{Wenwen Zhang}
\author[label3]{Steve Hankey}
\author[label3]{Meng Qi}
\author[label1]{Xiang Yan\corref{cor1}}

\address[label1]{Department of Civil and Coastal Engineering, University of Florida, Gainesville, FL 32603}
\address[label2]{Edward J. Bloustein School of Planning and Public Policy, Rutgers, the State University of New Jersey, New Brunswick, NJ 08901}
\address[label3]{School of Public and International Affairs, Virginia Tech, Blacksburg, VA 24061}

\ead{xiangyan@ufl.edu}

\cortext[cor1]{Corresponding author}

\begin{abstract}

\textcolor{black}{E-scooters have emerged as an important micromobility mode for short urban trips, yet evidence on route choice behavior remains limited. This study examines e-scooter route choice in Washington, DC using GPS trajectory data and a Path Size Logit model. In addition to roadway and infrastructure characteristics, the model incorporates visual streetscape features extracted from Google Street View imagery using computer vision techniques. The results show that the effectiveness of cycling infrastructure depends strongly on roadway context. On major roads, only protected bicycle facilities significantly increase route attractiveness, whereas on minor roads both protected and designated lanes provide utility gains. Sidewalks constitute the most frequently used riding environment, yet only asphalt-paved sidewalks are associated with positive utility, suggesting that sidewalk riding may reflect the absence of attractive on-street alternatives rather than a preference for pedestrian infrastructure. Tree coverage, particularly during summer, as well as building and wall coverage, are positively associated with route choice. Likelihood ratio tests and value-of-distance analysis indicate that roadway infrastructure exerts a stronger influence on route choice than visual streetscape features, although the latter provide additional explanatory power. These findings support targeted infrastructure investment and the integration of streetscape improvements as a complementary strategy for enhancing micromobility route attractiveness.}

\end{abstract}


\newpage

\end{frontmatter}

\section{Introduction}


\textcolor{black}{E-scooters have emerged as an important micromobility mode in cities worldwide, offering flexible and low-emission options for short urban trips. Understanding e-scooter route choice behavior is increasingly relevant for transportation planning, as it can guide the design of dedicated cycling facilities, improve rider safety, and support broader efforts toward sustainable urban mobility. While bicycle route choice has been extensively studied, e-scooters possess several characteristics that distinguish them from traditional bicycles, including higher-risk interactions with motor vehicles, unique physical properties such as small wheels, and distinct legal regulations governing their operation \citep{bielinski2020electric, badia2023shared, distefano2025comparison}. These differences suggest that route choice findings from bicycle contexts may not fully generalize to e-scooter travel, highlighting the need to develop an evidence base specific to e-scooter routing behavior. However, research on e-scooter route choice remains at an early stage. Existing studies have relied on stated preference methods \citep{huber2023scooter, sievert2023survey} or descriptive analyses of field experiments \citep{ringhand2024differences}. Only a limited number of studies have employed GPS-based revealed preference data to examine e-scooter route choice \citep{zhang2021type,cubells2023scooter, schumann2025city, chen2026modeling}, despite its ability to directly capture observed routing behavior \citep{fitch2020road, hsueh2023influential}. Among these, few have adopted a route-level discrete choice framework, which directly models the selection among route alternatives and yields utility-based estimates of attribute trade-offs. Given that route choice preferences vary substantially across urban contexts \citep{fitch2020road} and that existing findings are often bounded by local conditions \citep{cubells2023scooter}, additional GPS-based discrete choice evidence from diverse urban settings is needed to build a more robust understanding of e-scooter routing behavior.
}

\textcolor{black}{Beyond data and modeling approaches, the representation of route environment characteristics in existing e-scooter route choice studies remains relatively limited. Visual streetscape features have received limited attention in e-scooter route choice research; where included, coverage has been restricted to greenery \citep{cubells2023scooter, huber2023scooter}, despite broader evidence that visual environmental characteristics influence micromobility behavior \citep{juarez2023cyclists, he2024choose, xiao2025effects, lieu2026comparing}. Moreover, greenery has typically been treated as a homogeneous category, although it has been suggested that different vegetation types may exert distinct effects on route preferences \citep{xiao2025effects}. We also note that bicycle route choice research increasingly distinguishes not only among facility types but also among the roadway contexts in which they are provided, including differences in road hierarchy. However, such nuanced specifications have rarely been applied in e-scooter route choice modeling. In addition, sidewalk use is very common among e-scooter riders in the US, particularly in the absence of dedicated cycling facilities \citep{transportation2018scooter}, yet it is rarely examined explicitly in route choice models. Taken together, these limitations make it difficult to assess the relative influence of visual and physical route environments on e-scooter routing. While prior cycling studies suggest that infrastructure provision dominates route choice relative to streetscape features \citep{juarez2023cyclists}, whether this holds for e-scooter riders remains unclear, limiting the ability to prioritize alternative planning interventions.
}

\textcolor{black}{To address these gaps, this study examines e-scooter route choice behavior in Washington DC, a large US city where thousands of shared e-scooters have been deployed since 2017 and where substantial cycling infrastructure investment \citep{ddot2025bikelanes} coexists with a broader car-oriented urban environment. Leveraging shared e-scooter trip trajectory data from public APIs, a path size logit model is developed incorporating a detailed characterization of available cycling facilities. Visual streetscape features, including greenery classes and a broader set of scene elements, are additionally extracted from Google Street View imagery using computer vision techniques, complementing traditional GIS sources to provide a more comprehensive representation of the route environment. The study contributes new revealed preference evidence from a North American context on how physical and visual route attributes jointly shape e-scooter route choice.}

\textcolor{black}{The remainder of the paper is organized as follows. Section 2 reviews the literature on infrastructure and visual streetscape features in bicycle route choice, followed by current evidence on e-scooter route choice, including the data, modeling approaches, and research gaps. Section 3 describes the trip dataset, road network, and street view imagery. Section 4 presents the methodological framework, including model selection, choice set generation, and variable specification. Section 5 reports the model estimation results together with the Value-of-Distance (VoD) analysis. Section 6 discusses the findings in relation to infrastructure and streetscape planning, compares them with previous bicycle route choice research, and outlines their policy implications. Finally, Section 7 concludes the paper.}

\section{Literature Review}

\textcolor{black}{This section reviews the literature in three areas relevant to this study. Given the relative maturity of bicycle route choice research and the similarities between cycling and e-scooter travel environments, bicycle studies provide a useful foundation for understanding route choice behavior in micromobility contexts. We first review the influence of physical route attributes and cycling infrastructure, followed by visual streetscape features. We then examine the emerging evidence on e-scooter route choice to position the present study and identify remaining research gaps.}

\textcolor{black}{
\subsection{Infrastructure in bicycle route choice}
}

\textcolor{black}{A substantial body of bicycle route choice research has identified several route attributes that consistently influence cyclists' decisions. Cyclists generally prefer shorter routes and are sensitive to factors such as traffic signals, intersection complexity, traffic exposure, gradients, and turn frequency \citep{broach2012cyclists, fitch2020road, meister2024comparative}. Routes with frequent stops, steep slopes, or mixed traffic conditions are typically less favored, reflecting a preference for direct, comfortable, and uninterrupted travel.}

\textcolor{black}{Among these attributes, cycling infrastructure is a key determinant of route choice, and \citet{meister2023route} highlighted that infrastructure quality shapes route attractiveness. Cyclists generally prefer dedicated facilities over routes shared with motor vehicles \citep{meister2024comparative}. However, preferences vary across facility types and contexts: in Portland, off-street bike paths and bicycle boulevards show particularly strong positive effects \citep{broach2012cyclists}; in Copenhagen, protected bike lanes on major roads yield some of the largest utility gains \citep{fosgerau2023bikeability}, and painted bicycle lanes alone are found insufficient on large roads \citep{lukawska2023joint}; in Zurich, the relative preference for cycle lanes versus cycle paths varies across rider types \citep{meister2023route}. Furthermore, \citet{fitch2020road} found that in San Francisco, evidence that bicycling infrastructure matters for route choice is overwhelming, in contrast to Davis. Together, these studies suggest that the utility associated with cycling infrastructure varies substantially across facility types and the roadway contexts in which they are provided, and that route preferences may also differ across urban contexts.}

\textcolor{black}{
\subsection{Visual streetscape features in bicycle route choice}
}

\textcolor{black}{Visual streetscape features have been shown to influence a range of micromobility behaviors, including cycling experience, usage frequency, and route choice. Within cycling research, greenery has received the most attention: it enhances route attractiveness \citep{bialkova2022design}, increases usage and frequency \citep{gao2021urban, wang2020relationship}, and raises the likelihood of cycling \citep{lu2019associations, he2024choose}. In terms of bicycle route choice, cyclists generally favor greener routes \citep{park2019bicyclists, juarez2023cyclists, xiao2025effects, lieu2026comparing, lukawska2023joint}, attributed to pleasant scenery and comfort \citep{wang2020relationship} as well as reduced exposure to traffic, noise, and sunlight \citep{xiao2025effects}. However, it remains unclear whether these effects vary across vegetation types, as existing studies typically treat greenery as a homogeneous category \citep{xiao2025effects}. In addition, other visual attributes such as sky visibility and building enclosure have received comparatively less attention, though existing studies suggest their effects may be complex and context-dependent \citep{xiao2025effects, lieu2026comparing}. }

\textcolor{black}{Existing evidence also suggests that visual streetscape and physical infrastructure may play distinct roles in route choice. While infrastructure provision remains the dominant factor influencing cycling route choice \citep{juarez2023cyclists}, streetscape features primarily enhance route attractiveness rather than safety perception \citep{bialkova2022design}, and their effects on cycling propensity appear comparatively modest \citep{xiao2025effects}.}

\textcolor{black}{Recent advances in computer vision have expanded the capacity to measure visual streetscape features at scale. In the context of route choice modeling, street view imagery (SVI) offers two key advantages for capturing such attributes. First, matched to actual GPS-tracked paths, SVI enables precise measurement of riders' visual exposure along their genuine travel routes\citep{he2024choose}. Taking greenery as an example: eye-level greenness derived from SVI more accurately reflects cyclists' actual exposure and is more behaviorally relevant than overhead measures, which estimate greenspace in terms of park area or tree counts within a buffer zone \citep{lu2019associations, gao2021urban, wang2020relationship}. Second, SVI supports a broader and more fine-grained classification of streetscape elements: it enables the extraction of sky visibility, building enclosure, and other scene elements \citep{lieu2026comparing}, and allows greenery itself to be disaggregated into specific vegetation types \citep{helbich2019using}.}

\textcolor{black}{
\subsection{E-scooter route choice: current evidence and research gaps}
}

\textcolor{black}{
While e-scooters share some similarities with bicycles as a form of micromobility, they also exhibit several distinctive characteristics. Compared with bicycles, e-scooters attract a different user demographic, typically younger and higher-income individuals \citep{badia2023shared}, and are primarily used for shorter trips, with mean distances ranging from 1 to 4.7 km and durations between 7.6 and 20 minutes \citep{badia2023shared}. E-scooter riders also experience more frequent close interactions with motor vehicles and pedestrians than cyclists \citep{distefano2025comparison, transportation2018scooter}. Empirically, while e-scooter and bicycle route preferences show some similarities in terms of built environment responses, riders place different value on cycling lane types \citep{cubells2023scooter} and are more sensitive to road surface and infrastructure quality \citep{ringhand2024differences}. These differences suggest that findings from bicycle route choice research may not fully generalize to e-scooter travel. 
}

\textcolor{black}{
However, research on e-scooter route choice nevertheless remains limited. \citet{zhang2021type} conducted one of the earliest GPS-based studies using data from the Virginia Tech campus and applied a recursive logit model. \citet{huber2023scooter} employed a stated preference experiment analyzed using a conditional logit model, while \citet{cubells2023scooter} examined route detours in Barcelona using GPS trajectory data. More recently, \citet{chen2026modeling} modeled route choice in Brisbane using a Perturbed Utility Route Choice (PURC) model, and \citet{schumann2025city} applied a route-based discrete choice framework in Mannheim/Ludwigshafen, Germany. Additional evidence has been provided through field experiments \citep{ringhand2024differences} and survey-based studies \citep{sievert2023survey}.
}

\textcolor{black}{
These studies represent the emerging evidence base on e-scooter route choice, though they vary considerably in data type, modeling approach, and urban context. Regarding physical route attributes, a preference for road infrastructure is broadly consistent across existing studies and is further supported by e-scooter usage studies \citep{abouelela2023understanding, caspi2020spatial, qian2025big}. However, how e-scooter riders differentiate among specific facility types, particularly when provided in different roadway contexts, remains unclear. Furthermore, although previous e-scooter usage studies suggest that the influence of sidewalks is context-dependent \citep{abouelela2023understanding}, sidewalk use itself has rarely been explicitly examined in route choice studies. In North American contexts, bikeable sidewalks represent a common riding environment alongside on-street protected, painted, and shared bike lanes and off-street bike paths. Despite being discouraged or restricted in many jurisdictions \citep{ddot2025scooters}, sidewalk riding is frequently observed: in Portland, 1,622 sidewalk riding reports were submitted, representing 26.9\% of all reports \citep{transportation2018scooter}, and in a campus survey, \citet{zhang2021type} found sidewalks to be both the most preferred and most frequently used riding facility, with a usage rate of 70\%. Given the prevalence of sidewalk riding and ongoing concerns regarding pedestrian conflicts and safety, sidewalk use remains an important yet underexplored component of e-scooter route choice behavior.
}

\textcolor{black}{
Regarding visual streetscape features, street view imagery has rarely been incorporated into e-scooter route choice research, and the effects of visual elements such as sky, building, and wall on e-scooter routing remain largely unexplored. For greenery, existing evidence is limited and inconsistent. \citet{cubells2023scooter} found no significant detour behavior associated with urban greenness, \citet{huber2023scooter} reported mixed findings in descriptive analyses, while \citet{chen2026modeling} identified a positive effect of tree coverage. Moreover, existing studies have generally represented greenery using aggregate measures, leaving the influence of specific vegetation types insufficiently understood. Together, these findings highlight the need for a more comprehensive assessment of visual streetscape features and their role in shaping e-scooter route choice.
}

\textcolor{black}{
To address these gaps, this study applies a path size logit model to GPS-based e-scooter trajectory data from Washington DC. The model incorporates a detailed characterization of cycling facilities accessible to e-scooter riders, complemented by visual streetscape features extracted from Google Street View imagery using computer vision techniques. The resulting evidence provides new revealed preference insights into how physical and visual route attributes jointly shape e-scooter route choice in a large North American urban environment.
}

 
\section{Data}
\subsection{The GBFS and e-scooter trip trajectory data}
Washington DC was an early adopter of shared e-scooters, permitting seven operators, including Bird, Jump, Lime, Lyft, Razor, Skip, and Spin, to offer service beginning in 2017. The District Department of Transportation (DDOT) established operational requirements, including a mandate that vendors provide publicly accessible data through Application Programming Interfaces (APIs). These APIs deliver data in the General Bicycleshare Feed Specification (GBFS) format, a widely adopted open data standard for micromobility systems. GBFS data includes details such as vehicle ID, latitude and longitude, reservation and disabled status, and battery level. We developed a Python script to scrape GBFS data at one-minute intervals, although actual update frequencies vary by vendor (from one to ten minutes). The GBFS data recorded the GPS locations of shared e-scooters in the city over time. Because the GBFS data obtained from other vendors do not allow us to infer e-scooter trip trajectories or due to data quality issues, we decided to only keep Lime’s data for the period of February 21, 2019 to July 18, 2019 for route choice modeling, resulting in 125 consecutive days of observation.

Each entry in the dataset represents a GPS point, including latitude, longitude, a second-level timestamp, and a unique e-scooter ID that remains stable over time. This consistent ID allows us to accurately infer e-scooter trips and reconstruct the trip trajectories. In processing the data, we identified and removed periods during which a scooter repeatedly reported the same GPS coordinates at one-minute intervals over extended durations (e.g., one hour), indicating that the vehicle was stationary. These points do not reflect actual movement and may result from idling, GPS drift, or users' failure to end a trip.

\begin{figure}[!t]
  \centering
  \includegraphics[width=1\textwidth]{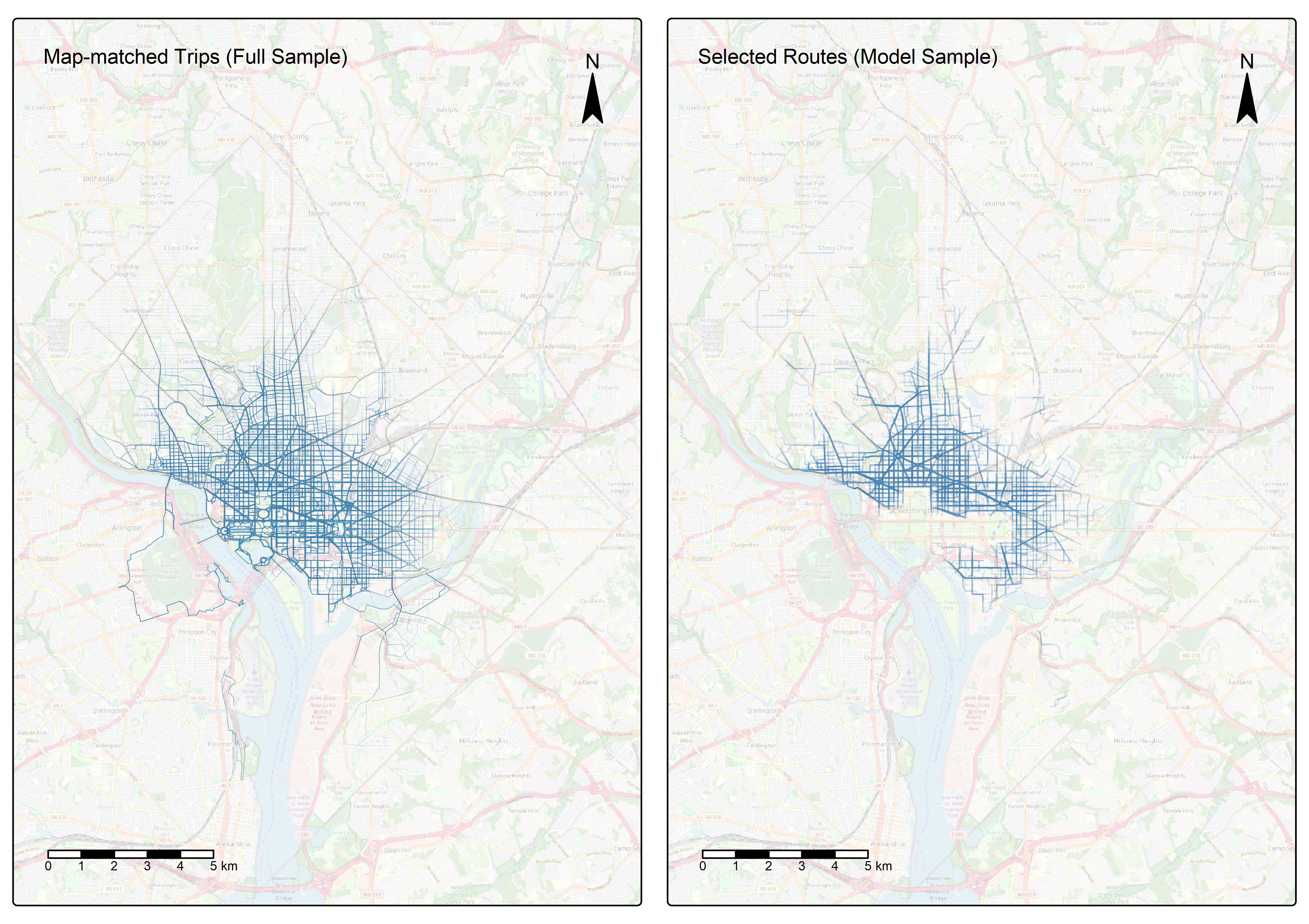}
  \caption{\textbf{Lime E-scooter Trips from February 21, 2019 to July 18, 2019}}\label{trips}
  \par\vspace{2mm}
  \raggedright
  \footnotesize\textbf{Note:} Trips near the DC National Park area were excluded, as such trips are likely recreational and may not reflect purposeful route choices.
\end{figure}

We then constructed trip trajectories by applying two criteria: (1) sequences of points must share the same ID and be sampled continuously, and (2) a gap of more than two minutes between consecutive points of the same ID was used to define trip start and end points. We then refined the trip data by excluding trips shorter than 2 minutes, longer than 90 minutes, with speeds outside 2–15 mph, or distances outside 0.4–3 miles. Additionally, we excluded all e-scooter trips near the DC National Park area, as such trips are likely recreational and may not reflect purposeful route choices, making them less suitable for modeling. After processing, 48,141 valid trips were identified for this study (Figure \ref{trips}), which incorporate all trip trajectories in DC.

\subsection{Network and map matching}
This section describes the preparation of the road network and GPS trajectory data for route choice modeling. It includes map matching, trajectory post-processing, and the identification of valid origin–destination (OD) pairs.

Map matching is a crucial step that assigns GPS-based trajectories to a series of links or nodes (and related network attributes) to precisely reconstruct users' actual paths. As reviewed in a study by \citet{berjisian2023evaluation}, map matching for active travel modes is more complex than for car travel, highlighting two main challenges: (1) the paths used by cyclists and pedestrians are not always represented in the street network; and (2) the complexity of active transportation networks, where multiple facilities may exist in parallel, such as bike lanes and sidewalks. Individuals often switch between closely parallel facilities that traditional matching algorithms struggle to detect. For e-scooters, these challenges are even more pronounced. E-scooter users have more flexible path choices, and their right of way on different road types is often intertwined and ambiguous.

We utilized Fast Map Matching (FMM) to process the trip trajectories, requiring not only the trajectory data but also the road network data \citep{yang2018fast}. OpenStreetMap (OSM) provides road network data compatible with FMM’s input requirements. The optional network levels as the FMM input include driving, biking, and walking. \textcolor{black}{Considering the ordinances on e-scooter road use in DC, e-scooters are permitted to use all designated bike lanes, are forbidden from riding on sidewalks in the central business district, and are encouraged to avoid sidewalks in pedestrian-heavy areas \citep{ddot2025scooters}}. Therefore, in practice, e-scooters are likely to use both bike lanes and sidewalks, depending on the specific location and context. Given that OSM’s walk-level network covers most of the bicycle-level network, we opted to use the walk-level network. This choice could introduce some inaccuracies, such as when an E-scooter actually travels on a link that is not represented in the walk-level network. However, such cases are very rare (5\%), and the resulting inaccuracies are considered acceptable. Conversely, using the less complicated bicycle-level network could have matched a significant portion of trips taken on sidewalks to bike lanes with significant offsets, creating a much larger error. \textcolor{black}{Another potential limitation of this approach is that, due to GPS positioning uncertainty, it is not always possible to reliably distinguish whether an e-scooter rider used a sidewalk or a parallel on-street bike lane when both facilities are present along the same corridor. Nevertheless, the map-matched trajectories generally remain spatially consistent with the actual routes taken, and any resulting ambiguity is limited to facility assignment on a small subset of segments.} Given the importance of network selection for subsequent route choice modeling, these considerations support the overall suitability of the walk-level network for the present study.

Using the selected valid trip trajectories and the walk-level network, we performed map matching (FMM), assigning each trip a unique ID. The output included the trip ID, the sequence of traversed links and nodes, and the corresponding geometries matched to the road network. We manually verified a subset of randomly sampled routes by comparing the map-matched paths to the original GPS trajectories. This examination revealed no significant issues, indicating the robustness of the used approach. 

To prepare inputs for the choice set generation algorithm, we conducted post-processing on the map-matching results. Trips containing duplicate nodes or links within a single trajectory—such as circular routes or back-and-forth movements—were excluded. These patterns often indicate non-purposeful or leisure travel rather than trips between distinct locations. In many such cases, the origin and destination are very close, resulting in unrealistically short alternative routes that do not reflect the user’s actual decision-making. Including these trips could compromise the validity of route choice modeling. Additionally, from the trajectory data, we extracted the sequence of nodes, with the first and last nodes designated as the origin and destination, respectively. For a trip with $n$ links, the corresponding trajectory contains $n-1$ confirmed traversed nodes. Therefore, a minimum of three links is required to define a valid OD pair. Trips consisting of only three or four links generally yield very small or no meaningful alternative route sets. Hence, only trips with at least five links were retained.

\subsection{Street View imagery processing and trip filtering}

This section describes the acquisition and processing of Google Street View (GSV) imagery, as well as the filtering of trips based on GSV coverage.

\textcolor{black}{
As discussed in the literature review, understanding travelers' route choices requires not only trip characteristics (e.g., travel time and distance) and available facilities, but also information about the visual environment experienced along the route. Traditional GIS-based measures, such as greenery indices, are typically derived from an overhead perspective and may not accurately represent riders' street-level visual exposure. Recent advances in computer vision have enabled the extraction of detailed streetscape features from Street View Imagery (SVI), providing a richer representation of the visual environment for route choice modeling. Given its extensive coverage in Washington DC, Google Street View (GSV) was used as the SVI source in this study.
}

We extracted visual streetscape features from GSV imagery using deep learning–based semantic segmentation. Images were processed using the Pyramid Scene Parsing Network (PSPNet), a state-of-the-art model that classifies each pixel into one of 150 categories based on the MIT Scene Parsing Benchmark and the ADE20K dataset \citep{zhou2019semantic}. These categories include natural elements (e.g., tree, plant, grass, sky) and built environment features (e.g., building, wall, pole). We computed the proportion of each feature at the image level to quantify streetscape composition. To ensure comprehensive spatial coverage, we used GSV images from across Washington, D.C., collected throughout 2019. Although the e-scooter data spans from February to July, the use of a full-year GSV dataset helps maintain temporal consistency while improving spatial completeness. The final dataset includes 10,854 unique image locations, each with associated latitude and longitude coordinates. \textcolor{black}{Each road segment was assigned the semantic features extracted from its nearest GSV sampling location. If the distance to the nearest GSV sampling location exceeded 50 m, the segment was assigned missing values.} Due to the sparse and uneven distribution of GSV points, only a subset of links could be enriched in this way. \textcolor{black}{To ensure adequate streetscape coverage, we retained only those trips for which at least 95\% of the route length was represented by road segments with valid streetscape feature values.} This filtering step significantly reduced missing data and improved the reliability of environmental feature estimates along each route. As a result, 2,057 trips were selected and their origin–destination pairs extracted for use in generating alternative routes.

\section{Methodology}
This section outlines the methodology used to model e-scooter route choice behavior. We begin by introducing the Path Size Logit (PSL) model, which accounts for overlapping routes and is well-suited for analyzing route choice behavior. We then describe the process of generating realistic and diverse route alternatives using the Breadth First Search Link Elimination (BFS-LE) algorithm. Finally, we detail the variables used in the model, including both traditional infrastructure attributes and streetscape features derived from imagery data.

\subsection{Path size logit model}

\textcolor{black}{
Route choice studies have employed both stated preference (SP) and revealed preference (RP) data. SP methods, typically based on surveys or recalled routes, can capture the preferences of both riders and non-riders, but may suffer from discrepancies between stated and actual behavior. GPS-based RP data, by contrast, directly capture observed routes and reduce this gap \citep{fitch2020road, hsueh2023influential}.
}

\textcolor{black}{
Within GPS-based route choice research, three broad modeling strategies can be distinguished. First, detour-based regression methods use detour percentage to explain deviations from the shortest path \citep{park2019bicyclists, cubells2023scooter}. Second, link-based methods, including the Recursive Logit \citep{fosgerau2013link} and Perturbed Utility Route Choice (PURC) \citep{fosgerau2022perturbed} models, assume that travelers make route decisions sequentially rather than selecting a complete route before departure, and therefore model route-building behavior at the link level. Third, route-based discrete choice methods compare complete route alternatives characterized by route-level attributes within a random utility maximization framework.
}

\textcolor{black}{
The Multinomial Logit (MNL) model is a common starting point for route-based analysis \citep{ben1985discrete}, but assumes independence among alternatives, making it unsuitable when routes share common segments \citep{bekhor2006evaluation, prato2009route}. The Path Size Logit (PSL) model addresses this limitation by incorporating a correction term that accounts for route overlap. Owing to this advantage, PSL has been widely applied in route choice research \citep{ton2018evaluating, marra2020determining, nielsen2021relevance, yap2021taking, sevtsuk2021big, meister2023route}. However, applications of route-based discrete choice models remain relatively rare in the e-scooter literature \citep{schumann2025city}. Given the objective of quantifying route-level trade-offs among competing alternatives, the PSL model is adopted in this study.
}

To calculate the path size term (\(PS_i\)) for route \(i\), we use the following formula:
\begin{equation}
PS_i = \sum_{a \in i} \frac{L_a}{L_i} \frac{1}{\sum_{j \in C} \delta_{aj}}
\end{equation}
where \(L_a\) is the length of link \(a\), \(L_i\) is the length of route \(i\), \(\delta_{aj}\) is an indicator function that equals 1 if link \(a\) is in route \(j\) and 0 otherwise, and \(C\) is the choice set of all routes.

The utility (\(U_i\)) for route \(i\) in the Path Size Logit (PSL) model is given by:
\begin{equation}
U_i = \beta X_i + \ln(PS_i)
\end{equation}
where \(U_i\) is the utility of route \(i\), \(\beta\) is a vector of estimated coefficients, \(X_i\) is a vector of observed attributes for route \(i\), and \(PS_i\) is the path size term for route \(i\).

The probability (\(P_i\)) of choosing route \(i\) is then given by the standard multinomial logit formula:
\begin{equation}
P_i = \frac{e^{U_i}}{\sum_{j \in C} e^{U_j}}
\end{equation}
where \(P_i\) is the probability of choosing route \(i\) and \(C\) is the choice set of all routes.

These equations together form the basis of the Path Size Logit model, which accounts for route overlap by incorporating the path size term into the utility calculation, thereby providing a more accurate representation of route choice behavior.

\subsection{Choice set generation}

\textcolor{black}{
The PSL model requires a choice set consisting of the observed route and a set of non-chosen alternatives, making choice set generation a critical methodological step. Most route generation methods rely on repeated shortest-path searches within the network. In this study, the Breadth First Search Link Elimination (BFS-LE) algorithm \citep{rieser2013route} was adopted due to its favorable balance between observed-route coverage and computational efficiency in dense active transportation networks \citep{halldorsdottir2014efficiency, ton2018evaluating}. Starting from the least-cost path for each OD pair, the algorithm iteratively removes links and recalculates shortest paths, generating alternative routes until the desired number of alternatives is reached or no additional valid paths can be identified.
}

\textcolor{black}{
Choice set quality is commonly evaluated based on observed-route coverage and within-set diversity. In the present study, the average coverage rate\footnote{Coverage rate here is defined as the proportion of the observed route length that is shared with a generated alternative.} of the highest-coverage alternative was 56\%, with 62\% of OD pairs containing at least one alternative exceeding 50\% coverage and 11\% containing an alternative exceeding 90\% coverage. These values are comparable to or slightly below those reported in bicycle route choice studies\footnote{\citet{felder2022choice} report that 11\% of OD pairs achieve 90\% coverage with five alternatives; \citet{fitch2020road} report 68\%--80\% of OD pairs achieve 50\% coverage across different cities.}, with the somewhat lower coverage likely reflecting the larger route choice space available to e-scooter riders in dense walk-level networks. To improve route heterogeneity while controlling choice set size, a diversification procedure similar to \citet{meister2023route} and \citet{tahlyan2020performance} was applied, whereby a newly generated route was retained only if its commonality factor with all previously accepted routes fell below 0.9, ensuring sufficient dissimilarity among alternatives. The mean within-OD standard deviation of coverage rates across alternatives is 16\%, confirming that the generated choice sets maintain meaningful diversity among alternatives. Up to five generated alternatives were retained per OD pair following \citet{felder2022choice}. The final dataset contains 2,057 OD pairs with an average of 5.5 routes per choice set, including the observed route.
}

\begin{ThreePartTable}
\begin{TableNotes}
\footnotesize
\item[a] ``Major'' refers to roads tagged in OSM as highway=\{``primary'', ``secondary'', ``tertiary'', ``unclassified''\} that have at least two car lanes in at least one direction.
\item[b] ``Minor'' refers to roads tagged in OSM as highway=\{``primary'', ``secondary'', ``tertiary'', ``unclassified''\} that have at most one car lane per direction.
\item[c] ``Designated'' refers to protected and painted bike lanes, which are combined because protected lanes were scarce on minor roads.
\end{TableNotes}

\renewcommand\arraystretch{1.5}
\footnotesize
\begin{longtable}{lp{11cm}}
\caption{\textcolor{black}{Variables considered in the e-scooter route choice model}}\label{var} \\
\hline
\textbf{Variable} & \textbf{Description} \\
\hline
\endfirsthead

\multicolumn{2}{l}{\footnotesize\textit{Table \ref{var} continued}} \\
\hline
\textbf{Variable} & \textbf{Description} \\
\hline
\endhead

\hline
\endfoot

\hline
\insertTableNotes
\endlastfoot

\multicolumn{2}{l}{\textbf{Trip characteristic}} \\
Route length (km)& Total length of all links in the route \\
Signals & Number of traffic signals along the route \\
\hline
\multicolumn{2}{l}{\textbf{Elevation}} \\
Slope 2\%-6\% (km)           & Total length of links with a positive 2\%-6\% slope \\
Slope 6\%-10\% (km)          & Total length of links with a positive 6\%-10\% slope \\
Slope \textgreater{}10\% (km)& Total length of links with a positive \textgreater{}10\% slope \\
\hline
\multicolumn{2}{l}{\textbf{Road infrastructure}} \\
Bike paths (km)             & Total length of links with OSM tag highway=``cycleway'' \\
\textcolor{black}{Sidewalks (asphalt) (km)}     & Total length of links with OSM tag highway=``footway'' with surface=``asphalt'' \\
\textcolor{black}{Sidewalks (non-asphalt) (km)} & Total length of links with OSM tag highway=``footway'' with surface other than ``asphalt'' or missing \\
Pedestrian zones (km)        & Total length of links with OSM tag highway=``pedestrian'' \\
Stairs (km)                  & Total length of links with OSM tag highway=``steps'' \\
Service (km)                 & Total length of links with OSM tag highway=``service'' \\
Major\textsuperscript{a} w/ protected (km)   & Length of major roads with protected bike lanes. OSM tag: cycleway=\{``track'', ``separate''\} \\
Major\textsuperscript{a} w/ painted (km)     & Length of major roads with painted bike lanes. OSM tag: cycleway=\{``lane'', ``opposite\_lane''\} \\
Major\textsuperscript{a} w/ shared (km)      & Length of major roads with shared bicycle markings. OSM tag: cycleway=\{``shared\_lane'', ``share\_busway''\} \\
Minor\textsuperscript{b} w/ designated\textsuperscript{c} (km) & Length of minor roads with designated bike lanes. OSM tag: cycleway=\{``lane'', ``opposite\_lane'', ``track'', ``separate''\} \\
Minor\textsuperscript{b} w/ shared (km)      & Length of minor roads with shared bicycle markings. OSM tag: cycleway=\{``shared\_lane'', ``share\_busway''\} \\
Residential (km)             & Total length of links with OSM tag highway=``residential'' \\
\hline
\multicolumn{2}{l}{\textbf{Streetscape feature}} \\
Tree               & Length-weighted average of tree pixel proportion along the route \\
Tree$\times$Summer & Interaction term between tree index and month (February--April = 0, May--July = 1) \\
\textcolor{black}{Plant}              & Length-weighted average of plant pixel proportion along the route \\
\textcolor{black}{Grass}              & Length-weighted average of grass pixel proportion along the route \\
Building           & Length-weighted average of building pixel proportion along the route \\
Wall               & Length-weighted average of wall pixel proportion along the route \\
Sky                & Length-weighted average of sky pixel proportion along the route \\
Pole               & Length-weighted average of pole pixel proportion along the route \\
\hline
Path size & Path size term \\

\end{longtable}
\end{ThreePartTable}

\subsection{Variables included in e-scooter route choice model}

\begin{figure}[!t]
  \centering
  \includegraphics[width=0.7\textwidth]{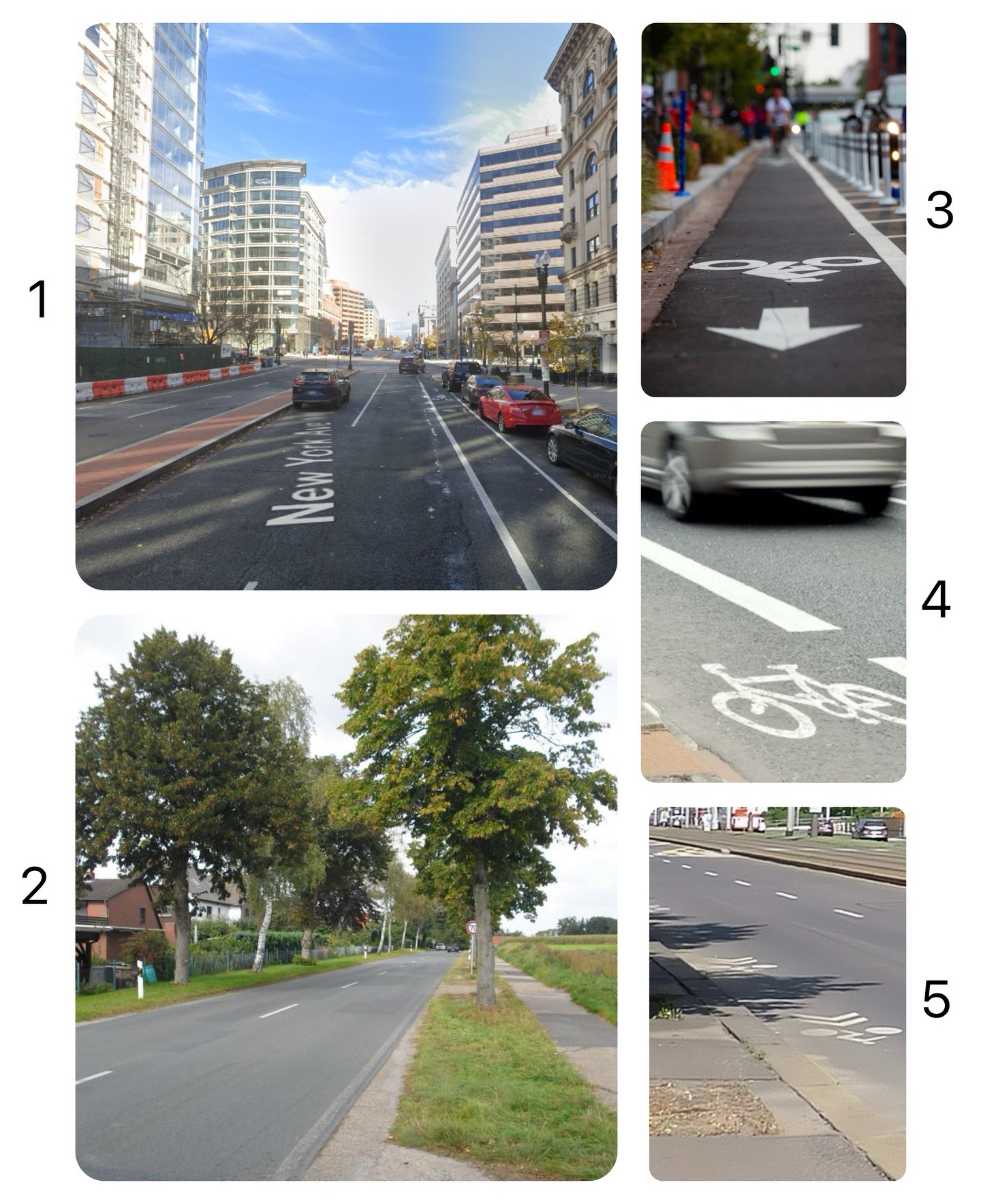}
  \caption{Illustration of roadway characterization and bicycle facility types: 
  (1) major roads, 
  (2) minor roads, 
  (3) protected lanes, 
  (4) painted lanes, 
  (5) shared lanes.
  Source: OpenStreetMap (OSM), Google street view, District Department of Transportation (DDOT).}
  \label{var1}
\end{figure}

To identify the factors that may influence e-scooter users' route choices, we generated attributes for all routes within the choice set. We have considered four categories of variables: trip characteristics, elevation, road infrastructure, and streetscape features, Table \ref{var} provides detailed information about these variables. Figure \ref{var1} illustrates representative roadway and facility types.

All elevation and road infrastructure variables were measured in kilometers. For infrastructure variables, we summed the lengths of links tagged with specific facility types in OSM. In the literature on micromobility and bicycle route choice, road environments are commonly represented using either the original OSM road classifications or infrastructure-oriented classifications derived from combinations of OSM tags. The former primarily captures roadway functional class rather than the cycling facilities available to riders. As a result, it is difficult to isolate the effects of specific cycling infrastructure within broad roadway categories such as primary, secondary, and tertiary roads. Following \citet{lukawska2023joint}, we therefore transformed the original OSM road classifications into an infrastructure-oriented specification by combining "highway=*" and "cycleway=*" tags. For example, a segment tagged as "highway=primary" with "cycleway=separate" was classified as a protected bicycle facility on a primary road. \textcolor{black}{To reduce the number of explanatory variables while preserving roadway context, main road classes were further aggregated according to the number of car lanes, as explained in Table \ref{var}. This classification enables examination of how specific cycling facility types vary across different roadway contexts while keeping the number of explanatory variables manageable. In addition to cycling facility and roadway context, pavement surface quality has also been identified as an important determinant of e-scooter route choice \citep{sievert2023survey, huber2023scooter, ringhand2024differences, chen2026modeling}. In the study area, on-street bicycle facilities are overwhelmingly asphalt-paved. By contrast, sidewalk surfaces exhibit substantially greater variation and are therefore classified into asphalt and non-asphalt categories.}

Streetscape features were aggregated to the route level using a link length–weighted average. Elevation attributes were computed using 2’ contour elevation data from Open Data DC. We estimated slope by extracting the closest elevation points to the origin and destination of each segment and dividing the elevation difference by segment length. \textcolor{black}{Based on prior literature \citep{meister2023route}}, slopes were categorized into ranges: 2\%–6\%, 6\%–10\%, and greater than 10\%. 

\textcolor{black}{The model was estimated using a long-format dataset with a route-level binary dependent variable. Within each OD-specific choice set, the observed route was coded as 1 and all alternative routes were coded as 0.}

\begin{table}[!t]
\centering
\footnotesize
\renewcommand\arraystretch{1.2}
\caption{Comparison of attribute values of selected routes and alternatives}\label{descriptive}
\vspace{4pt}
\resizebox{\textwidth}{!}{%
\begin{tabular}{lcccccc}
\hline
& \multicolumn{2}{c}{Selected routes} & & \multicolumn{2}{c}{Alternative routes} & \\
\cmidrule(lr){2-3}\cmidrule(lr){5-6}
Variable & Mean (SD) & Proportion & & Mean (SD) & Proportion & Diff. \\
\hline
\multicolumn{7}{l}{\textbf{Trip characteristic}} \\
Route length (m)  & 912 (276) & 100\%  & & 923 (272) & 100\%  & 0.00\% \\
Signals           & 3.3 (2.4)     & /      & & 3.4 (2.5)     & /      & /      \\
\hline
\multicolumn{7}{l}{\textbf{Elevation}} \\
Slope 2\%-6\% (m)        & 151 (122) & 16.57\% & & 152 (122) & 16.50\% & 0.07\%  \\
Slope 6\%-10\% (m)       & 67 (78)   & 7.33\%  & & 67 (79)   & 7.26\%  & 0.07\%  \\
Slope $>$10\% (m)        & 73 (102)  & 8.04\%  & & 77 (104)  & 8.33\%  & -0.29\% \\
\hline
\multicolumn{7}{l}{\textbf{Road infrastructure}} \\
Bike paths (m)              & 193 (301) & 21.11\% & & 176 (283) & 19.02\% & 2.09\%  \\
Sidewalks (asphalt) (m)      & 44 (60)   & 4.78\%  & & 41 (55)   & 4.39\%  & 0.39\%  \\
Sidewalks (non-asphalt) (m)  & 372 (276) & 40.74\% & & 364 (269) & 39.45\% & 1.29\%  \\
Pedestrian zones (m)        & 1 (11)    & 0.13\%  & & 1 (10)    & 0.10\%  & 0.03\%  \\
Stairs (m)                  & 1 (10)    & 0.09\%  & & 3 (19)    & 0.32\%  & -0.23\% \\
Service (m)                 & 15 (45)   & 1.62\%  & & 24 (61)   & 2.57\%  & -0.95\% \\
Major w/ protected (m)      & 58 (146)  & 6.31\%  & & 56 (145)  & 6.03\%  & 0.28\%  \\
Major w/ painted (m)        & 20 (73)   & 2.15\%  & & 22 (76)   & 2.36\%  & -0.21\% \\
Major w/ shared (m)         & 12 (64)   & 1.36\%  & & 14 (72)   & 1.57\%  & -0.21\% \\
Minor w/ designated (m)     & 102 (180) & 11.23\% & & 97 (179)  & 10.53\% & 0.70\%  \\
Minor w/ shared (m)         & 11 (52)   & 1.22\%  & & 11 (51)   & 1.18\%  & 0.04\%  \\
Residential (m)             & 72 (166)  & 7.90\%  & & 66 (148)  & 7.16\%  & 0.74\%  \\
\hline
\multicolumn{7}{l}{\textbf{Streetscape feature}} \\
Tree        & 16.64\% (7.79\%) & / & & 16.54\% (7.54\%) & / & 0.10\% \\
Plant       & 1.34\% (1.74\%)  & / & & 1.34\% (1.70\%)  & / & 0.00\% \\
Grass       & 1.03\% (1.48\%)  & / & & 1.05\% (1.53\%)  & / &  -0.03\%\\
Building    & 17.76\% (8.74\%) & / & & 17.70\% (8.52\%) & / &  0.06\%\\
Wall        & 1.27\% (3.34\%)  & / & & 1.26\% (3.38\%)  & / &  0.01\%\\
Sky         & 18.22\% (5.75\%) & / & & 18.29\% (5.66\%) & / &  -0.07\%\\
Pole        & 0.03\% (0.05\%)  & / & & 0.03\% (0.05\%)  & / &  0.00\%\\
\hline
\end{tabular}%
}
\par\vspace{2mm}
\footnotesize\textbf{Note:} Proportion refers to the share of route length. SD = standard deviation.
\end{table}

\section{Results}
\subsection{Descriptive statistics for chosen vs. alternative routes}

We first compared the attribute values of the selected routes and those of the non-chosen alternatives to 1) help form hypotheses of the direction of impacts; 2) verify the validity of choice set generation results. Table \ref{descriptive} presents the mean and variance of each route attribute. It was observed that the mean length of the selected routes is greater. This is expected because the choice set comprises the shortest path and slightly deviated second shortest paths. In fact, we found that approximately 20\% of trips chose the shortest path. 

To reflect the preference for elevation and road infrastructure relative to the shortest paths, we normalized the total length of both sets to 100\% and then calculated the proportion of distance-based variables. We found that the proportions for slope 2\%-6\%, bike paths, sidewalks, residential roads, major roads with protected bike lanes, and median roads with designated bike lanes were positive, suggesting a positive effect. Conversely, the proportions of major roads with shared lanes, stairs, and service roads, as well as slopes exceeding 10\% were negative, indicating a potential negative effect. Streetscape-related statistics are similar between selected and alternative routes. However, this does not necessarily imply that these factors have no influence on route choice decisions; their effects will be further examined in the route choice model.

\subsection{Model outputs}

\textcolor{black}{
Table \ref{results} presents the model estimates. Four models were estimated to assess the incremental explanatory contribution of different variable groups. Model 1 includes only trip characteristics and elevation variables as a baseline. Models 2 and 3 extend Model 1 by separately adding GSV-derived streetscape features and GIS-derived infrastructure variables, respectively. Model 4 is the full model incorporating all variable groups. The full model achieves a Concordance Index of 0.72, indicating good ability to rank the observed chosen route above non-chosen alternatives. The inclusion of streetscape variables improves McFadden $R^2$ from 0.16 to 0.17, a modest but statistically significant gain confirmed by a likelihood ratio test comparing Model 4 against Model 3 (p$<$0.001). This increment is comparable to that reported in bicycle route choice studies incorporating streetscape features: \citet{xiao2025effects} report an improvement from 0.19 to 0.20, and \citet{lieu2026comparing} from 0.17 to 0.19. Importantly, infrastructure variables contribute greater incremental explanatory power than streetscape variables, as reflected in the larger improvement in model fit from Model 1 to Model 3 relative to Model 1 to Model 2, consistent with prior evidence that physical infrastructure dominates route choice relative to visual environment \citep{juarez2023cyclists}. Coefficient estimates remain largely consistent in both sign and magnitude across model specifications, indicating a stable underlying preference structure. Accordingly, the following discussion focuses on the full model (Model 4). The positive coefficient on the path size term is expected and consistent with findings from BFS-LE-based PSL models \citep{meister2024comparative}, reflecting the degree of overlap inherent in the generated choice sets.}

In the PSL model, route length is negatively and highly significantly associated with route choice, indicating that e-scooter users prefer shorter routes when other variables are held constant. This finding is consistent with most existing bicycle and e-scooter route choice studies \citep{halldorsdottir2014efficiency,meister2024comparative,lukawska2023joint,meister2023route,ton2018evaluating,scott2021route}. In Table \ref{VoD_infra}, the coefficient of Route Length is summed with those of other distance-based road attributes to obtain their total effects. All total effects are negative, indicating that although certain road attributes positively influence route choice, the additional travel distance results in an overall negative effect. This finding supports the validity of the model. Since recreational trips were excluded so that the remaining trips primarily reflect daily travel needs, such as commuting and shopping, this result further underscores the decisive role of travel distance in route choice.

\begin{table}[!t]
\centering
\footnotesize
\renewcommand\arraystretch{1.2}
\caption{\textcolor{black}{E-scooter route choice model results}}\label{results}
\vspace{4pt}
\resizebox{\textwidth}{!}{%
\begin{tabular}{lcccccccc}
\hline
& \multicolumn{2}{c}{Model 1 (Base)} 
& \multicolumn{2}{c}{Model 2 (GSV)} 
& \multicolumn{2}{c}{Model 3 (Infra)} 
& \multicolumn{2}{c}{Model 4 (Full)} \\
\cmidrule(lr){2-3}\cmidrule(lr){4-5}\cmidrule(lr){6-7}\cmidrule(lr){8-9}
Variable & Coef. (SE) & z & Coef. (SE) & z & Coef. (SE) & z & Coef. (SE) & z \\
\hline
\multicolumn{9}{l}{\textbf{Trip characteristic}} \\
Route length   & $-4.48^{***}$ (0.52) & $-8.59$ & $-4.55^{***}$ (0.53) & $-8.62$ & $-4.69^{***}$ (0.58) & $-8.13$ & $-4.72^{***}$ (0.58) & $-8.08$ \\
Signals        & $0.02$ (0.02)         & $1.38$  & $0.02$ (0.02)         & $1.40$  & $0.03$ (0.03)         & $1.19$  & $0.03$ (0.03)         & $1.08$  \\
\hline
\multicolumn{9}{l}{\textbf{Elevation}} \\
Slope 2\%-6\%         & $0.25$ (0.55) & $0.45$ & $0.14$ (0.55) & $0.26$ & $0.14$ (0.56) & $0.25$ & $0.00$ (0.57) & $0.00$ \\
Slope 6\%-10\%        & $1.18$ (0.84) & $1.41$ & $1.17$ (0.85) & $1.38$ & $1.49^{.}$ (0.87) & $1.73$ & $1.46^{.}$ (0.88) & $1.66$ \\
Slope $>$10\%         & $0.33$ (1.05) & $0.32$ & $0.13$ (1.08) & $0.12$ & $0.44$ (1.10) & $0.40$ & $0.24$ (1.13) & $0.22$ \\
\hline
\multicolumn{9}{l}{\textbf{Road infrastructure}} \\
Bike paths            &               &        &               &        & $0.51^{*}$ (0.23)     & $2.21$ & $0.51^{*}$ (0.23)     & $2.19$ \\
Sidewalks (asphalt)     &               &        &               &        & $1.96^{**}$ (0.73)    & $2.67$ & $1.97^{**}$ (0.74)    & $2.68$ \\
Sidewalks (non-asphalt) &               &        &               &        & $0.36$ (0.23)         & $1.55$ & $0.32$ (0.24)         & $1.34$ \\
Pedestrian zones      &               &        &               &        & $5.61$ (3.72)         & $1.51$ & $6.50^{.}$ (3.78)     & $1.72$ \\
Stairs                &               &        &               &        & $-20.40^{***}$ (4.95) & $-4.12$& $-22.10^{***}$ (5.01) & $-4.42$\\
Service               &               &        &               &        & $-4.30^{***}$ (0.80)  & $-5.37$& $-4.47^{***}$ (0.81)  & $-5.53$\\
Major w/ protected    &               &        &               &        & $0.81^{*}$ (0.34)     & $2.42$ & $0.77^{*}$ (0.34)     & $2.28$ \\
Major w/ painted      &               &        &               &        & $0.02$ (0.61)         & $0.03$ & $-0.01$ (0.61)        & $-0.01$\\
Major w/ shared       &               &        &               &        & $0.50$ (0.68)         & $0.73$ & $0.54$ (0.68)         & $0.79$ \\
Minor w/ designated   &               &        &               &        & $1.17^{***}$ (0.29)   & $4.06$ & $1.14^{***}$ (0.29)   & $3.92$ \\
Minor w/ shared       &               &        &               &        & $0.75$ (0.90)         & $0.83$ & $0.63$ (0.91)         & $0.69$ \\
Residential           &               &        &               &        & $1.19^{***}$ (0.32)   & $3.71$ & $1.16^{***}$ (0.33)   & $3.56$ \\
\hline
\multicolumn{9}{l}{\textbf{Streetscape feature}} \\
Tree                  &               &        & $6.43^{*}$ (2.98)     & $2.16$ &               &        & $7.55^{*}$ (3.11)     & $2.43$ \\
Tree$\times$Summer    &               &        & $20.60^{*}$ (9.80)    & $2.10$ &               &        & $20.08^{*}$ (10.00)   & $2.01$ \\
Plant                 &               &        & $2.89$ (6.90)        & $0.42$&               &        & $2.47$ (7.14)    & $0.35$\\
Grass                 &               &        & $-5.47$ (5.10)        & $-1.07$&               &        & $-8.70^{.}$ (5.29)    & $-1.65$\\
Building              &               &        & $6.71^{*}$ (2.73)     & $2.46$ &               &        & $7.26^{*}$ (2.85)     & $2.55$ \\
Wall                  &               &        & $8.13^{**}$ (3.09)    & $2.63$ &               &        & $9.99^{**}$ (3.38)    & $2.96$ \\
Sky                   &               &        & $5.06$ (3.27)         & $1.55$ &               &        & $5.94^{.}$ (3.41)     & $1.74$ \\
Pole                  &               &        & $-299.68^{*}$ (150.14)& $-2.00$&               &        & $-388.18^{*}$ (159.72)& $-2.43$\\
\hline
Path Size             & $2.38^{***}$ (0.09) & $27.47$ & $2.37^{***}$ (0.09) & $27.19$ & $2.36^{***}$ (0.09) & $26.48$ & $2.35^{***}$ (0.09) & $26.19$ \\
\hline
\multicolumn{9}{l}{\textit{Model fit}} \\
$n$                   & \multicolumn{2}{c}{11384} & \multicolumn{2}{c}{11384} & \multicolumn{2}{c}{11384} & \multicolumn{2}{c}{11384} \\
Events                & \multicolumn{2}{c}{2057} & \multicolumn{2}{c}{2057} & \multicolumn{2}{c}{2057} & \multicolumn{2}{c}{2057} \\
McFadden $R^2$        & \multicolumn{2}{c}{0.14} & \multicolumn{2}{c}{0.15} & \multicolumn{2}{c}{0.16} & \multicolumn{2}{c}{0.17} \\
Concordance           & \multicolumn{2}{c}{0.697}& \multicolumn{2}{c}{0.701}& \multicolumn{2}{c}{0.713}& \multicolumn{2}{c}{0.715}\\
LRT (vs Null)         & \multicolumn{2}{c}{$862.88^{***}$} & \multicolumn{2}{c}{$885.95^{***}$} & \multicolumn{2}{c}{$993.47^{***}$} & \multicolumn{2}{c}{$1023.17^{***}$} \\
LRT (vs Base)         & \multicolumn{2}{c}{---}  & \multicolumn{2}{c}{$23.07^{**}$} & \multicolumn{2}{c}{$130.60^{***}$} & \multicolumn{2}{c}{$160.29^{***}$} \\
LRT (vs Infra)        & \multicolumn{2}{c}{---}  & \multicolumn{2}{c}{---}  & \multicolumn{2}{c}{---}  & \multicolumn{2}{c}{$29.70^{***}$} \\
\hline
\end{tabular}}
\par\vspace{2mm}
\footnotesize\textbf{Note:} $^{***}$ $p<0.001$, $^{**}$ $p<0.01$, $^{*}$ $p<0.05$, $^{.}$ $p<0.1$
\end{table}

We found that e-scooter route choice is not sensitive to signal variables, which is somewhat unexpected. One possible explanation is that the high maneuverability of e-scooters allows riders to adapt more easily to signalized environments. Rather than avoiding signals, riders may prioritize route directness and convenience, accepting signal delays when necessary. Existing studies report mixed findings. Consistent with our results, \citet{zhang2021type} found the number of traffic signals to be insignificant in a campus setting. \textcolor{black}{In contrast, \citet{huber2023scooter} reported a negative effect of intersection density, while \citet{cubells2023scooter} found that e-scooter riders avoid signalized intersections but prefer routes with higher crosswalk density. Together, these findings suggest that the effect of traffic signals is context-dependent and may interact with other intersection characteristics.}

In terms of slope, gradients of 2\%–6\% have a positive coefficient that is marginally significant, while other slope categories are not significant. This differs from most bicycle route choice studies, where the effect of slope becomes increasingly negative as the gradient increases. However, our results are consistent with \citet{meister2023route}, who found that e-bikes tolerate steeper uphill gradients than conventional bicycles because electric assistance offsets much of the disutility of climbing. Similarly, \citet{zhang2021type} reported that e-scooter users are not sensitive to road slopes. The marginally positive effect of 2\%–6\% slopes may reflect the greater flexibility of e-scooter users in route selection. In addition, an alternative measure based on the variance of elevation change also showed a positive effect, suggesting that e-scooter users may prefer routes with certain attributes rather than simply avoiding slopes \citep{scott2021route}. Overall, these findings suggest that slope is not a major constraint on e-scooter route choice, with other route attributes playing a more important role.

\textcolor{black}{We found that e-scooter users prefer asphalt sidewalks and pedestrian zones. The positive effect of asphalt sidewalks, together with the insignificant effect of non-asphalt sidewalks, is consistent with previous studies showing that e-scooter riders are sensitive to pavement quality \citep{chen2026modeling, ringhand2024differences, sievert2023survey}. Because e-scooters have small wheels and limited suspension, rough surfaces reduce riding comfort and increase perceived instability \citep{huber2023scooter}.The high proportion of non-asphalt sidewalks in selected routes (Table \ref{descriptive}), together with these findings, may also suggest that sidewalk use is often driven by necessity rather than preference.} E-scooter users prefer local roads, such as residential streets, likely because of their lower traffic volumes and speeds. In contrast, stairs are strongly avoided, consistent with \citet{zhang2021type}, as they require riders to dismount and carry the scooter or negotiate unsafe conditions. \textcolor{black}{Service roads also have a negative effect, likely because they mainly provide access to adjacent properties rather than continuous travel corridors.}

The estimation results reveal heterogeneous effects across bicycle facility types (protected, painted, and shared) and roadway contexts (major vs. minor roads). On major roads, only protected lanes are positively and significantly associated with route choice, whereas painted lanes and shared markings are not significant, suggesting that only protected facilities provide sufficient separation in high-traffic environments. On minor roads, the combined category of designated lanes (protected and painted) is strongly significant. These findings indicate that the utility gap between protected and painted facilities is context dependent: pronounced on major roads but minimal on minor roads, where lower traffic volumes and narrower roadways may allow even painted lanes to provide sufficient comfort. In contrast, shared lanes show no significant effect in either roadway context, suggesting a limited role in promoting e-scooter use. \textcolor{black}{These findings are broadly consistent with the emerging e-scooter literature, suggesting that riders do not value all bicycle facilities equally \citep{huber2023scooter, sievert2023survey, chen2026modeling}, and align with bicycle route choice research showing that the benefits of cycling infrastructure vary substantially by facility type and roadway context \citep{lukawska2023joint, fosgerau2023bikeability}.}

\textcolor{black}{
Among the streetscape variables, tree, building, wall, and pole show significant effects. Tree coverage is positively associated with route choice, with the effect further strengthened during summer through an interaction with season. This is consistent with previous bicycle route choice studies \citep{park2019bicyclists, xiao2025effects, lieu2026comparing} and likely reflects the benefits of shade and thermal comfort. In contrast, plants have no significant effect, while grass is marginally negative, suggesting that riders respond to the functional qualities of vegetation rather than greenery itself. Sky visibility exhibits a marginal positive association with route choice. Previous studies have reported complex and context-dependent effects of sky visibility, potentially reflecting its relationship with road width, urban form, and local environmental conditions \citep{xiao2025effects}. Building coverage and wall coverage are positively associated with route choice, possibly reflecting riders' preference for more urbanized and active street environments. Pole coverage is negatively associated with route choice. Because poles in street view imagery encompass a broad category of vertical infrastructure, including traffic signal poles, street lights, and utility poles, higher pole coverage may reflect more visually complex and traffic-intensive environments that are less attractive to e-scooter riders.}

\begin{table}[!t]
\centering
\renewcommand\arraystretch{1.2}
\caption{Value-of-Distance indicators: Road infrastructure}\label{VoD_infra}
\vspace{4pt}
\begin{tabularx}{\textwidth}{Xrrr}
\hline
Variable & Marginal Effect & Total Effect & VoD \\
\hline
\multicolumn{4}{l}{\textbf{Road infrastructure}} \\
Sidewalks (asphalt)    & $1.97$  & $-2.75$ & $-0.42$ \\
Residential            & $1.16$  & $-3.56$ & $-0.25$ \\
Minor w/ designated    & $1.14$  & $-3.58$ & $-0.24$ \\
Major w/ protected     & $0.77$  & $-3.95$ & $-0.16$ \\
Bike paths             & $0.51$  & $-4.21$ & $-0.11$ \\
Service                & $-4.47$ & $-9.19$ & $0.95$ \\
Stairs                 & $-22.10$ & $-26.82$ & $4.68$ \\
\hline
\end{tabularx}
\par\vspace{2mm}
\raggedright
\small\textbf{Note:} Only variables with $p < 0.05$ are reported. Marginal Effect is the coefficient of the variable. Total Effect is the sum of the variable coefficient and the route length coefficient ($-4.72$). VoD is the ratio of the variable coefficient to the route length coefficient.
\end{table}

\subsection{Value-of-Distance: infrastructure effects}
In the context of route choice models, the Value of Distance (VoD) metrics are commonly calculated. These metrics represent the equivalent distance that would be traded off for a particular attribute along the route. For linear parameter models, VoDs are given by the ratio of parameter i to the length parameter. Table \ref{VoD_infra} presents the marginal effects, total effects, and VoD indicators for various variables. VoDs above or below 0 are considered to increase or decrease the perceived length by the corresponding VoD ratio. It is important to note that the negative utility of distance cannot be fully offset by any single road attribute, and thus cannot make an overall positive contribution to utility.

\textcolor{black}{
The negative effects of certain route features are particularly pronounced for stairs and service roads. On routes with stairs, the length parameter decreases by 22.10, indicating that e-scooter riders are willing to travel 468\% extra distance to avoid them. Service roads also show a strong negative total effect of -4.47, with a VoD of 95\%, consistent with their role as access ways for buildings, parking lots, and service facilities rather than continuous travel corridors. Conversely, several facility types reduce perceived route distance. Major roads with protected bike lanes, median roads with designated bike lanes, and residential roads have total effects of -3.95, -3.58, and -3.56, with VoD indicators of -16\%, -24\%, and -25\%, respectively, indicating that riders view these as safer and more continuous options. Off-street bike paths reduce perceived distance by approximately -11\%, a more modest gain, likely reflecting their limited connectivity with the overall transportation network. Surface quality also plays an important role in shaping route utility. Asphalt-paved sidewalks show the strongest VoD among all facility-related variables at -42\%, suggesting that pavement quality may be as influential as facility provision in shaping e-scooter route preferences.}

\textcolor{black}{
\subsection{Value-of-Distance: streetscape effects}
}

\textcolor{black}{
Unlike road infrastructure variables, streetscape features are measured as pixel proportions (0--100\%) rather than distances. Consequently, their VoD values represent the effect of a hypothetical change from 0\% to 100\% coverage and are not directly comparable to the VoD of infrastructure variables. A more meaningful interpretation considers marginal changes based on observed variation. As shown in Table \ref{VoD_street}, a one standard deviation increase in tree coverage (approximately 7.79\%) corresponds to a perceived distance reduction of around 125 m in spring and 456 m in summer. For wall and building coverage, the equivalent reductions are 71 m and 134 m per standard deviation, respectively. These results indicate that streetscape features can meaningfully influence perceived route attractiveness. However, the feasible variation in streetscape characteristics within a given OD pair is generally limited, as evidenced by the relatively small differences between selected and alternative routes in Table \ref{descriptive}. Consequently, although visual streetscape features contribute to route preferences, their practical influence remains more limited than that of major infrastructure characteristics. This finding is consistent with the likelihood-ratio-test results and further supports the conclusion that distance and infrastructure define the foundation of route selection, while streetscape features primarily modify route attractiveness at the margin.
}

\begin{table}[!t]
\centering
\renewcommand\arraystretch{1.2}
\caption{Value-of-Distance indicators: Streetscape features}\label{VoD_street}
\vspace{4pt}
\begin{tabularx}{\textwidth}{Xrrrr}
\hline
Variable & Marginal Effect & VoD (km) & SD (\%) & Per SD effect (m)\\
\hline
Tree (spring)  & $7.55$     & $-1.60$ & $7.79$ & $-125$ \\
Tree (summer)  & $27.63$    & $-5.85$ & $7.79$ & $-456$ \\
Wall           & $9.99$     & $-2.12$ & $3.34$ & $-71$  \\
Building       & $7.26$     & $-1.54$ & $8.74$ & $-134$ \\
Pole           & $-388.18$  & $82.24$ & $0.05$ & $+41$  \\
\hline
\end{tabularx}
\par\vspace{2mm}
\raggedright
\small\textbf{Note:} Only variables with $p < 0.05$ are reported. VoD (km) is the perceived distance change for a full 0--100\% change in pixel proportion. SD (\%) is the standard deviation among selected routes. Per SD effect reports the equivalent perceived distance change in metres for a one standard deviation increase.
\end{table}

\textcolor{black}{Finally, two points regarding the VoD estimates warrant clarification. First, although the proposed framework enables direct comparison between streetscape features and cycling facilities, this does not imply that they are interchangeable in practice. For example, increased tree coverage may enhance route attractiveness but is unlikely to substitute for high-quality cycling infrastructure, particularly in terms of rider safety. These attributes likely influence route choice through different mechanisms, and visual environment improvements alone are unlikely to achieve comparable safety benefits \citep{bialkova2022design}. Second, VoD estimates are model dependent. \citet{meister2024comparative} reported substantial variation in the magnitudes and rankings of bike lane VoDs across model specifications using the same dataset. For example, the Recursive Logit model was generally less sensitive to route attributes than BFS-LE-based PSL models. Coefficient estimates may also be influenced by model specification and multicollinearity, although all VIF values in our model were below 3. Policy decisions should therefore place greater emphasis on the sign and relative ranking of VoD indicators than on their exact magnitudes, and ideally be informed by evidence from multiple studies.}

\section{Discussion}

\textcolor{black}{
The discussion focuses on three main themes: the role of roadway infrastructure and sidewalk use in shaping route preferences; the contribution of visual streetscape features to route attractiveness; and the similarities and differences between e-scooter and bicycle route choice behavior.}

\subsection{Roadway infrastructure and sidewalk use}

On major roads, only protected bicycle facilities significantly increase route attractiveness, whereas on lower-volume minor roads both protected and designated lanes provide utility gains. \textcolor{black}{This indicates that the effectiveness of a facility depends on the surrounding traffic environment.} Shared lane markings, however, do not significantly influence route choice on either major or minor roads, \textcolor{black}{suggesting that symbolic designation alone provides limited benefit to e-scooter riders and may be insufficient to encourage micromobility use.} By contrast, residential streets are strongly preferred, even where dedicated bicycle facilities are absent. This may reflect their lower traffic volumes and speeds, which can provide a relatively comfortable riding environment. \textcolor{black}{Together, these findings imply that infrastructure investment should be tailored to roadway context.}

\textcolor{black}{
A notable finding concerns sidewalk use. Sidewalks constitute the most frequently used riding environment in our sample, exceeding the combined use of on-street bicycle facilities. This is consistent with evidence that sidewalk riding increases when dedicated cycling infrastructure is absent \citep{transportation2018scooter}. However, only asphalt-paved sidewalks show a positive effect, whereas the predominantly non-asphalt sidewalks provide no significant utility gain. These findings imply that frequent sidewalk riding may be driven more by the lack of attractive on-street alternatives than by a genuine preference for pedestrian infrastructure, raising concerns about rider safety and pedestrian conflicts. Particular attention should be given to major roads where riders currently rely on sidewalks because of the lack of dedicated cycling facilities. In these locations, expanding on-street bicycle infrastructure should remain the preferred long-term solution, while improving sidewalk surface quality may serve as a practical complementary measure where dedicated cycling infrastructure is difficult to provide.
}

\subsection{Role of visual streetscape features}
Our analysis also demonstrates the role of visual streetscape features in shaping e-scooter route preferences. The positive effect of tree coverage likely reflects the functional benefits of shade and thermal comfort rather than aesthetic appeal alone. \textcolor{black}{The contrasting effects of different vegetation types further support this interpretation, suggesting that riders respond to the functional qualities of vegetation rather than greenery in general.} \textcolor{black}{The positive associations with buildings, walls, and sky visibility further indicate that route attractiveness is influenced by multiple aspects of the street-level environment.}

\textcolor{black}{
The likelihood ratio tests and VoD analysis consistently indicate that roadway infrastructure has a stronger influence on route choice than visual streetscape features, although the latter remain statistically significant and provide additional explanatory power. These findings suggest that streetscape and infrastructure play complementary rather than substitutable roles in shaping e-scooter route preferences, contributing to the ongoing discussion of their relative importance \citep{juarez2023cyclists, bialkova2022design, xiao2025effects, park2019bicyclists}. Infrastructure likely influences route choice primarily through safety and riding conditions, whereas streetscape features contribute to route attractiveness and comfort. Consequently, while streetscape improvements cannot substitute for dedicated cycling infrastructure, they may provide a practical strategy. Compared with major infrastructure reconstruction, such improvements can often be implemented incrementally \citep{wang2020relationship}.}

\subsection{Comparisons with Bicycle Route Choice}

E-scooter users and cyclists share broadly similar route choice patterns, with some notable differences. The central role of bicycle infrastructure in shaping e-scooter route choice, together with the positive effect of tree coverage, is consistent with findings from bicycle route choice research, suggesting that cycling studies provide a useful baseline for understanding e-scooter behavior. However, e-scooter users show greater tolerance for slopes, likely because of electric assistance. In addition, they make greater use of sidewalks and appear to be more sensitive to sidewalk pavement quality than conventional cyclists. These findings indicate that, although many insights from bicycle route choice research transfer to e-scooter travel, mode-specific characteristics continue to shape route preferences.

\section{Conclusion}
This study examined e-scooter route choice in Washington DC using a path size logit model applied to GPS trajectory data from shared e-scooters. The results show that on-street bicycle infrastructure plays a central role in shaping route preferences, with its effectiveness varying by roadway context: protected facilities are critical on major high-volume roads, while residential streets provide comfortable riding conditions even without dedicated infrastructure. E-scooter riders also exhibit sensitivity to pavement surface quality. Beyond infrastructure, visual streetscape features contribute meaningfully to route attractiveness. Tree coverage exerts a positive effect on route choice, further amplified in summer, while building and wall coverage are also positively associated with route utility. VoD analysis indicates that stairs impose the largest disutility, while asphalt sidewalks provide the greatest utility gains among the infrastructure variables. Among the streetscape variables, tree coverage in summer produces the largest utility gain.

\textcolor{black}{
These findings carry practical planning implications. Dedicated cycling infrastructure should remain the primary strategy for improving e-scooter travel conditions, particularly on major roads where protected facilities are absent and sidewalk riding is consequently more common. Where dedicated facilities are difficult to provide, improving sidewalk surface quality may offer a practical complementary measure. Streetscape improvements, particularly increased tree coverage, can further enhance route attractiveness and comfort. Together, these findings suggest that integrating high-quality cycling infrastructure with supportive streetscape design can create safer and more attractive micromobility environments.}

Several limitations should be acknowledged. First, the absence of demographic and socioeconomic variables means the model assumes uniform preferences across the population; future research should examine preference heterogeneity by incorporating e-scooter rider characteristics. \textcolor{black}{Second, streetscape features were extracted using a general-purpose semantic segmentation model applied to street view imagery. In addition, because street view images are collected from vehicle-mounted cameras, they may not perfectly represent the visual environment experienced by riders. Third, the data cover only February to July 2019, limiting the temporal representativeness of the findings. Seasonal variation in streetscape preferences, particularly during autumn and winter months when canopy cover is reduced and thermal comfort concerns differ, cannot be assessed with the current dataset.} Finally, this study focuses on Washington DC. The estimated relationships may differ in cities with different climates, urban forms, infrastructure provision, and regulatory environments, highlighting the need for future validation across multiple urban contexts.

\bibliographystyle{elsarticle-harv}
\biboptions{semicolon,round,sort,authoryear}
\bibliography{sample}

\end{document}